\documentclass[showkeys,showpacs,preprintnumbers,amsmath,amssymb]{revtex4}
\usepackage{amsmath}
\usepackage{txfonts}
\usepackage{color}
\usepackage[colorlinks,linkcolor=red,anchorcolor=blue,citecolor=green]{hyperref}
\usepackage{graphicx}
\usepackage{rotating}
\usepackage{dcolumn}
\usepackage{bm}

\begin{document}

\title{From one to infinity: symmetries of integrable systems}
\author{S. Y. Lou and Man Jia}
\affiliation{\footnotesize{School of Physical Science and Technology, Ningbo University, Ningbo, 315211, China}}

\date{\today}

\begin{abstract}
Integrable systems constitute an essential part of modern physics. Traditionally, to approve a model is integrable one has to find its infinitely many symmetries or conserved quantities. In this letter, taking the well known Korteweg-de Vries and Boussinesq equations as examples, we show  that it is enough to find only one nonlocal key-symmetry to guarantee the integrability. Starting from the nonlocal key-symmetry, recursion operator(s) and then infinitely many symmetries and Lax pairs can be successfully found.
\end{abstract}
\maketitle
Usually, a nonlinear system is called integrable, one has to point out that the model is integrable under what special meanings. A Lax integrable model possesses a Lax pair such that the model can be considered as a consistent condition of the Lax pair \cite{Lax}. A Painlev\'e integrable model requires all of the movable singularities of its all solutions with respect to an arbitrary singular manifold are poles \cite{Pain,Pain1,Pain2,Pain3}. An IST integrable model is solvable by means of the inverse scattering transformation \cite{IST}. A CRE (or CTE) integrable model can be solved by means of the consistent Riccati expansion (or the consistent Tanh expansion) method \cite{CRE,Bose,CTE}. A symmetry integrable system is defined as it possesses infinitely many symmetries \cite{Bluman, Olver}. To find infinitely many symmetries is clearly not easy work. For (1+1)-dimensional nonlinear systems, a fundamental method is to find a recursion operator such that infinitely many symmetries can be obtained by repeatedly applying the recursion operator on some trivial seed symmetries like the travelling symmetries, Galilean invariance invariance and scaling invariance \cite{RO,RO1}. For (2+1)-dimensional systems, one can use the so-called mastersymmetry approach \cite{master} and the formal series symmetry approach \cite{formal} to find infinitely many symmetries.

Here, we propose a significant question: can we find one key symmetry to guarantee the integrability for a nonlinear system? In other words, can we find infinitely many symmetries from only one symmetry?

The study of symmetries is fundamental to find or establish universal models like the standard model \cite{Higgs} in particle physics. There are some types of symmetry methods to solve complicated nonlinear physical problems. The symmetry approach is more attractive in the study of integrable systems because the existence of infinitely many local and nonlocal symmetries \cite{Olver}. Local symmetries are widely used to get symmetry invariant solutions, to reduce the dimensions of partial differential equations and to find new integrable systems. Recently, it was found that nonlocal symmetries are also very useful to find novel types of
exact solutions \cite{LouPLB,LouJPA,LouPRE,LouPRE1,Bose}, integrable models, and the relations among different types of integrable hierarchies \cite{LouNon,lou2021JHEP}. It is worth to mention that combining some local and nonlocal symmetries one can find interaction solutions among different types of nonlinear waves including solitary waves, cnoidal periodic waves, Bessel waves, Airy waves, rational waves, Painlev\'e waves \cite{LouJPA,LouPRE,LouPRE1,LouNon,lou2021JHEP,Bose}, Korteweg-de Vries (KdV) waves and Boussinesq waves \cite{Wave}. Basing on the results of the local-nonlocal symmetry reduction method, one may propose some more general approaches like the consistent Riccati (or tanh function) expansion method to find more general interaction solutions \cite{CRE,Bose,CTE}.

The function $\sigma$ is called a symmetry of the evolution equation
\begin{eqnarray}
u_t=K(u),
\end{eqnarray}
if we have always
\begin{eqnarray}
\frac{\textrm{d}\sigma}{\textrm{d} t}=K'\sigma,
\end{eqnarray}
where $K'$ is the linearized operator of $u$. The symmetry means the evolution equation is invariant under the transformation $u\rightarrow u+\epsilon \sigma$ with infinitesimal parameter $\epsilon$. A recursion operator, strong symmetry is defined if it satisfies
\begin{eqnarray}
\frac{\textrm{d} \Theta}{\textrm{d}t}=[K', \Theta]=K' \Theta-\Theta K'. \label{ro}
\end{eqnarray}
In other words, if $\sigma$ is a symmetry of an evolution equation, then $\Theta \sigma$ with $\Theta$ being given by \eqref{ro} is also a symmetry of the same equation.

In this letter, we point out that for a nonlinear physical model, it may be enough to find a key nonlocal symmetry to guarantee its integrability. That means we can find recursion operator(s) and then infinitely many symmetries from one nonlocal symmetry.

To realize this idea, we first take the well known KdV equation,
\begin{equation}
u_t=6uu_x+u_{xxx},\label{KdV}
\end{equation}
as a simple example. The KdV equation applies to a very large variety of physical fields, such as nonlinear optics, Bose-Einstein condensates, hydrodynamics, acoustics, plasma physics, solid state physics, gravity, biology, and many other areas \cite{KdV1,KdV2,KdV3,KdV4}.

A symmetry, $\sigma$, of the KdV equation \eqref{KdV} is defined as a solution of the linearized equation of \eqref{KdV}
\begin{equation}
\sigma_t=6\sigma u_x+6u\sigma_x+\sigma_{xxx},\qquad K'=6u_x+6 u \partial_x +\partial_x^3,\label{sym}
\end{equation}
which means the KdV equation \eqref{KdV} is invariant under the transformation $u\rightarrow u+\epsilon \sigma$ with infinitesimal parameter $\epsilon$.

It is straightforward to check that
\begin{equation}
u=-\frac{2f_x^2}{f^2}+2\frac{u_1}{f}+u_2,\label{NBT}
\end{equation}
with
\begin{eqnarray}
u_1&=&f_{xx},\label{u1}\\
u_2&=&\lambda-\frac12\frac{f_{xxx}}{f_x}+\frac14\frac{f_{xx}^2}{f_x^2},\label{u2}
\end{eqnarray}
and
\begin{equation}
f_t=6\lambda f_x+f_{xxx}-\frac{3}{2}\frac{f_{xx}^2}{f_x}\label{ft}
\end{equation}
solves the KdV equation \eqref{KdV}. It is also interesting that $u_2$ expressed by \eqref{u2} with \eqref{ft} also solves the KdV equation \eqref{KdV}. That means Eq. \eqref{NBT} is an auto-B\"acklund transformation which transforms one solution $u_2$ to another solution $u$ for the same KdV equation. And Eq. \eqref{u2} is instead a nonauto-B\"acklund transformation which transforms the solution $f$ of the Schwartz KdV equation \eqref{ft} to the solution $u$ of another equation, the KdV equation \eqref{KdV}. The results \eqref{NBT}-\eqref{ft} have been derived in \cite{Pain1} by using the Painlev\'{e} property analysis.

Another important fact is that $u_1$ defined by \eqref{u1} is a nonlocal symmetry, residual symmetry \cite{Bose}, of the solution $u_2$ given by \eqref{u2}. One can directly check that $\sigma=f_{xx}$ satisfies the symmetry equation \eqref{sym} with the solution $u$ given by \eqref{u2}.

Now, an important question is can we derive some other integrable properties such as the recursion operator(s), infinitely many symmetries and Lax pair(s) from the residual symmetry?

To derive the recursion operator(s) of the KdV equation \eqref{KdV} from the single residual symmetry
\begin{eqnarray}
\sigma=f_{xx},\label{residual}
\end{eqnarray}
we rewrite \eqref{u2} as
\begin{eqnarray}
&& 2u {f_x}+f_{xxx}=2\lambda{f_x}+\frac12\frac{f_{xx}^2}{f_x},\label{ru2}
\end{eqnarray}
where $u_2$ has been redenoted as $u$ for simplicity.

Differentiating Eq. \eqref{ru2} with respect to $x$, we have
\begin{eqnarray}
&& 2u_x f_x+2u f_{xx}+f_{xxxx}=2\lambda f_{xx}+\frac{f_{xx}f_{xxx}}{f_x} -\frac12\frac{f_{xx}^3}{f_x^2}.\label{du2}
\end{eqnarray}
By eliminating $f_{xxx}$ on the right hand side of \eqref{du2} with help of \eqref{ru2}, one can immediately find
\begin{eqnarray}
&&\Phi \sigma =4\lambda \sigma,\qquad \Phi\equiv \partial_x^2+4u +2u_x\partial_x^{-1}, \label{Phi}
\end{eqnarray}
where the function $f$ has been cancelled by the residual symmetry \eqref{residual}. Here the operator $\partial^{-1}$ is the inverse operator of $\partial$, i.e., the partial integration operator satisfying $\partial \partial^{-1}=\partial^{-1} \partial=1$. Obviously, $\Phi$ is just the recursion operator of the KdV equation \eqref{KdV} and Eq. \eqref{Phi} is nothing but the eigenvalue problem of the recursion operator $\Phi$. The recursion operator $\Phi$ given by \eqref{Phi} has been found in many literatures.

Whence a recursion operator is obtained, the infinitely many symmetries are followed immediately by repeated applying the recursion operator on any other seed symmetries such as $u_x$ and $3tu_x+\frac12$ related to the space translation and the Galilean transformation, respectively.

Using the similar approach to the Schwarz KdV equation \eqref{ft} and the residual symmetry \eqref{residual}, we have another eigenvalue problem
\begin{eqnarray}
&&\Psi \sigma =-4\lambda \sigma,\qquad \Psi\equiv (\partial_t-2u\partial_x +2u_x)\partial_x^{-1}. \label{Psi}
\end{eqnarray}
The eigenvalue problem \eqref{Psi} can also be obtained from the symmetry equation \eqref{sym} by cancelling $\sigma_{xx}$ via \eqref{Phi}. It is not difficult to show that the two-dimensional operator $\Psi$ defined in \eqref{Psi} is also a recursion operator of the KdV equation \eqref{KdV} and the 2-dimensional operator $\Psi$ defined in \eqref{Psi} has not yet appeared before in literature.

The KdV hierarchy can be written as
\begin{eqnarray}
u_{t_{2n+1}}=\Phi^n u_x,\ n=0,\ 1,\ 2,\ \ldots \label{KdVh}
\end{eqnarray}
and/or
\begin{eqnarray}
u_{t_{2n+1}}=\Psi^n u_x,\ n=0,\ 1,\ 2,\ \ldots. \label{KdVh1}
\end{eqnarray}
The equivalence of \eqref{KdVh} and \eqref{KdVh1} can be proven by cancelling $u_t$ in \eqref{KdVh1} via the KdV equation \eqref{KdV}.

It is straightforward to verify that the compatibility condition of the eigenvalue problem \eqref{Phi} (or \eqref{Psi}) and the symmetry equation \eqref{sym} is just the KdV equation \eqref{KdV}. In other words, the linear equation system \eqref{Phi} and \eqref{sym} is a Lax pair of the KdV equation \eqref{KdV}. It is known that the square eigenfunction symmetry
\begin{equation}
\sigma =(\psi^2)_x \label{square}
\end{equation}
transforms the symmetry equations, the Lax pair \eqref{Phi} and \eqref{sym} to the traditional known Lax pair
\begin{eqnarray}
&&\begin{array}{l}\label{Lax}
\displaystyle{\psi_{xx}+u\psi=\lambda \psi,}\\ \\
\displaystyle{\psi_t=4\psi_{xxx}+6u\psi_x+3u_x\psi.}
\end{array}
\end{eqnarray}
Thus, we see that the existence of the nonlocal residual symmetry for the KdV equation is equivalent to the integrability of the KdV equation because the recursion operator(s) and then the infinitely many symmetries, conservation laws and Lax pairs can be derived from the single residual symmetry.
In fact, the N-fold Darboux transformations of the KdV equation can also be derived from the residual symmetry \cite{Tang} and/or the square eigenfunction symmetry \cite{LouCPL}.
The recursion operator of the KdV equation can also be derived from the square eigenfunction symmetry \eqref{square} and the infinitesimal nonlocal symmetry comes from the B\"acklund transformation \cite{LouJPA}.

To provide further support for our idea, we consider another well known integrable system, the Boussinesq equation \cite{Bq1,Bq2,Bq3,Bq4}
\begin{equation}
u_{tt}=\frac13(u_{xx}+4u^2)_{xx}, \label{BQ}
\end{equation}
which can be equivalently rewritten as
\begin{eqnarray}
&&\begin{array}{l}\label{Bq}
\displaystyle{u_t=v_x,}\\ \\
\displaystyle{v_t=\frac13 (u_{xx}+4u^2)_x.}
\end{array}
\end{eqnarray}
The Boussinesq equation \eqref{BQ} was introduced in 1871 for the propagation of long surface waves on water of constant depth \cite{Bq1,Bq2}.

Similar to the KdV case, we have an auto-B\"{a}cklund transformation
\begin{eqnarray}
&&\begin{array}{l}\label{aBq}
\displaystyle{u'=-\frac32\frac{f_x^2}{f^2}+\frac32\frac{f_{xx}}{f}+u,}\\ \\
\displaystyle{v'=-\frac32\frac{f_xf_t}{f^2}+\frac32\frac{f_{xt}}{f}+v}
\end{array}
\end{eqnarray}
with a nonauto-B\"{a}cklund transformation
\begin{eqnarray}
&&\begin{array}{l}\label{naBq}
\displaystyle{u=\frac38\frac{f_t^2+f_{xx}^2}{f_x^2}-\frac12\frac{f_{xxx}}{f_x},}\\ \\
\displaystyle{v=\lambda-\frac12\frac{f_{xt}}{f_x}+\frac14\frac{(f_tf_{xx})_x}{f_x^2}
-\frac14\frac{(f_t^2+f_{xx}^2)f_t}{f_x^3}},
\end{array}
\end{eqnarray}
where $\{u, v\}$ is a solution of \eqref{Bq}, $\{u', v'\}$ is a solution of \eqref{aBq} and $f$ is a solution of the Schwartz Boussinesq equation
\begin{eqnarray}
&&\begin{array}{l}\label{SBq}
\displaystyle{f_{tt}+f_{xxxx}=\frac{f_{xx}}{f_x^2}\big(f_t^2+4f_xf_{xxx}-3f_{xx}^2\big).}\
\end{array}
\end{eqnarray}
Naturally, the coefficients of $f^{-1}$ of \eqref{aBq},
\begin{eqnarray}\label{BqSym}
&&\sigma=\left(\begin{array}{l}
\displaystyle{\sigma^u}\\
\displaystyle{\sigma^v}
\end{array}\right)=\left(\begin{array}{l}
\displaystyle{f_{xx}}\\
\displaystyle{f_{xt}}
\end{array}\right),
\end{eqnarray}
where a trivial constant factor $3/2$ has been dropped out, is just a nonlocal symmetry, the residual symmetry of the Bossinesq equation \eqref{Bq}. In other words, \eqref{BqSym} solves
\begin{eqnarray}\label{Bqsy}
\left(\begin{array}{l}
\sigma^u\\
\sigma^v\end{array}\right)_t=\left(\begin{array}{cc}
0 & \partial_x \\
\frac13 (\partial^3_{x}+8\partial_x u) & 0
\end{array}\right)\left(\begin{array}{l}
\sigma^u\\ \sigma^v
\end{array}\right).
\end{eqnarray}
After finishing some simple calculations from the nonauto-B\"acklund transformation \eqref{naBq}, one can find that the following two relations
\begin{eqnarray}\label{fx3tfx6}
&&\begin{array}{l}
\displaystyle{f_{xxxt}+3 v f_{xx}+2v_x f_x+2 u f_{xt}+u_xf_t-3f_{xx}\lambda=0,}\\ \\
\displaystyle{\frac13f_{xxxxxx}+\frac{10}3 u f_{xxxx}+5 u_x f_{xxx}+3 u_{xx} f_{xx}
+\frac{16}3u^2f_{xx}}\displaystyle{
+\frac23f_x u_{xxx}+\frac{16}3f_xuu_x+3vf_{xt}+v_xf_t-3\lambda f_{xt}=0,}
\end{array}
\end{eqnarray}
are identically satisfied by considering the Schwartz Boussinesq equation \eqref{SBq}.

Using the residual symmetry condition \eqref{BqSym}, the equation system \eqref{fx3tfx6} is just the eigenvalue problem
\begin{equation}\label{Eign}
\Phi \left(\begin{array}{c}
\sigma^u\\ \sigma^v
\end{array}\right) =3\lambda \left(\begin{array}{c}
\sigma^u\\ \sigma^v
\end{array}\right)
\end{equation}
of the recursion operator
\begin{equation}\label{PhiBq}
\Phi = \left(\begin{array}{cc}
3 v +2v_x \partial_x^{-1} & \partial_{x}^2+2 u+u_x\partial^{-1}_x \\ \Phi_{21} & 3v+v_x\partial^{-1}_x
\end{array}\right)
\end{equation}
with $\Phi_{21}\equiv\frac13\partial^4_{x}+\frac{10}3 u \partial^2_{x}+5 u_x \partial_{x}+3 u_{xx}
+\frac{16}3u^2
+\frac23 (u_{xx}+4u^2)_x \partial_x^{-1}$.

Obviously, the eigenvalue problem \eqref{Eign} and the symmetry equation \eqref{Bqsy} constitute a special Lax pair of the Boussinesq equation \eqref{Bq}. Thus, we have derived the recursion operator and then the infinitely many symmetries and Lax pair of the Boussinesq equation \eqref{Bq} from the single nonlocal residual symmetry \eqref{BqSym}.

If one directly study the residual symmetry of the Boussinesq equation \eqref{BQ} instead of \eqref{Bq}, then the nonauto-B\"acklund transformation is described by the first equation of \eqref{naBq} with \eqref{SBq}. Starting from the first equation of \eqref{naBq}, one can find that
the following related linear system with respect to $f$
\begin{equation} \label{Psi1}
f_{xxxt}+3  f_{xx} (\partial_x^{-1}u_t-\lambda)+2u_t f_x+2 u f_{xt}+u_xf_t=0
\end{equation}
is satisfied. Because $f_{xx}$ is a nonlocal residual symmetry of the Boussinesq equation \eqref{BQ},
Eq. \eqref{Psi1} becomes an eigenvalue problem of the recursion operator $\Psi$
\begin{eqnarray} \label{PsiSy}
&& \Psi\sigma=3\lambda\sigma,\\
&&\Psi\equiv
\partial_{xt}+3 \left(\partial_x^{-1}u_t\right)+2u_t \partial^{-1}_x+2 u \partial_x^{-1}\partial_{t}+u_x\partial_{x}^{-2}\partial_t.\label{Psisy}
\end{eqnarray}
The eigenvalue problem \eqref{PsiSy} and the linearized equation of the Boussinesq equation \eqref{BQ}
\begin{equation}
\sigma_{tt}=\frac13(\sigma_{xx}+8u\sigma)_{xx}, \label{BQsy}
\end{equation}
constitute a special Lax pair of \eqref{BQsy}. The recursion operator $\Psi$ has not yet reported before.

Though the recursion operators $\Phi$ and $\Psi$ defined in \eqref{PhiBq} and \eqref{Psisy} are different at first glance, the Boussinesq hierarchies
\begin{eqnarray} \label{Bqh1}
&&\left( \begin{array}{l}
u \\ v
\end{array}\right)_{t_{2n+1}}=\Phi^n \left( \begin{array}{l}
u \\ v
\end{array}\right)_x, \ \left( \begin{array}{l}
u \\ v
\end{array}\right)_{t_{2n+2}}=\Phi^n \left( \begin{array}{l}
v \\ \frac{u}3+\frac43u^2
\end{array}\right)_x,
\end{eqnarray}
and
\begin{eqnarray} \label{Bqh2}
&&u_{t_{2n+1}}=\Psi^n u_x, \ u_{t_{2n+2}}=\Psi^n u_t,\ n=0,\ 1,\ 2,\ \ldots,
\end{eqnarray}
are equivalent. The equivalence can be directly proved by using the relations \eqref{Bq}.

In summary, for a (1+1)-dimensional nonlinear system, if one can find one key symmetry such that the recursion operator(s) and infinitely many symmetries and Lax pair(s) can be successfully obtained, then we say that the model is integrable under the meaning that it possesses a key symmetry. For simplicity, we say the system is key-symmetry-integrable (KSI). In this letter, we have proven that both the KdV equation and the Boussinesq system are KSI models. The key-symmetries used here are the nonlocal residual symmetries which can be obtained simply by using the truncated Painlev\'e analysis \cite{Pain1,Pain2}. In fact, one can check that some other types of nonlocal symmetries can be used as key-symmetries such as the square eigenfunction symmetries \cite{JMP94,JNMP94}, infinitesimal B\"acklund transformations \cite{LouJPA}, nonlocal symmetries related to the CRE/CTE approach \cite{CRE}.

It is generally accepted the different concepts of integrability are often equivalent. For instance, the Painlev\'{e} integrable implies the model possesses a Lax pair. While to solve a system by using IST, it is required to find the Lax pair of the system. The KSI means the system possesses the recursion operator(s) and Lax pair(s), thus, the KSI is equivalent to the Lax integrable in some sense. However, the KSI may not be equivalent to the Painlev\'{e} integrable in two aspects. One is some non-Painlev\'{e} integrable systems may possess a key symmetry. The prototypical example is the Harry-Dym equation. Although the well known Harry-Dym equation has nonlocal symmetries related to the recursion operator \cite{Lou1996} that can be treated as key-symmetries, it has been shown the equation is a weak Painlev\'{e} equation and has no Painlev\'{e} property \cite{Pain2}. The other is the KSI requires less conditions compared with the Painlev\'{e} integrable. The KdV equation is Painlev\'{e} integrable with five conditions by using the WTC method \cite{Pain1}, while the KSI requires only one condition because the singularity manifolds conditions have been changed to the Schwarz KdV equation.

These results may help us to further understand a fundamental problem, what is integrability? By now, it is more reasonable to say a general nonlinear system is integrable under some special meanings.

For a general unknown model, searching for the key symmetries and Lax pair(s) is not easy despite various effective methods for finding symmetries and constructing Lax pairs have been provided. Many types of nonlocal symmetries can be considered as the key symmetries, it is still difficult to determine the key symmetry related to the recursion operator(s) of the model because little results about the recursion operators have been known. Fortunately, once the key symmetry is found, the Lax pair(s) for the model can also be derived. Our results may provide a direct method to construct the Lax pair(s) for a model.

It is also interesting that the nonlocal key symmetries possess various other elegant properties in addition to derive recursion operator(s), infinitely many symmetries and Lax pairs. These types of nonlocal key symmetries can be localized to find many types of interaction solutions among different types of nonlinear excitations \cite{LouJPA,LouPRE,LouPRE1,Wave,Bose}. Algebro-geometric solutions can be obtained from the nonlinearization approach \cite{Cao} via nonlocal key-symmetry constraints \cite{LiCheng,Kono}. Applying nonlocal key-symmetry constraints on the Lax pairs, various kinds of new integrable systems can be obtained \cite{LouHu97}. The nonlocal key-symmetries can be used as sources to describe the interactions between long waves and short waves \cite{Source,Source1,Source2}. The localization of the nonlocal key-symmetries to the Camassa-Holm type systems \cite{CH}, the reciprocal transformations will be naturally included \cite{LouNon}.

In higher dimensions, for some types of dispersionless integrable systems like the Hirota equations and heavenly equations \cite{Heaven,Heaven1,Heaven2,Heaven3}, the same idea can be used to find the recursion operators, infinitely many symmetries and Lax pairs starting from suitable nonlocal key-symmetry. For (2+1)-dimensional dispersive nonlinear systems like the Kadomtsev-Petviashvili (KP) equation, B-type KP equation, Davey-Stewartson equation and Nizhnik-Novikov-Veselov equation, nonlocal key-symmetries (such as the square eigenfunction symmetries and residual symmetries) are really existent. However, how to use these nonlocal key-symmetries to find the other infinitely many symmetries is still an open problem.

\section*{Acknowledgement}
The author are in debt to the helpful discussions with Professors X. B. Hu and Q. P. Liu.
The authors acknowledge the support of the National Natural Science Foundation of China (Nos. 12275144, 12235007 and 11975131) and K. C. Wong Magna Fund in Ningbo University.

\providecommand{\newblock}{}

\end{document}